\documentstyle[aps,prl,floats]{revtex}

\newcommand{\bastar}{\begin{eqnarray*}}
\newcommand{\eastar}{\end{eqnarray*}}
\newskip\humongous \humongous=0pt plus 1000pt minus 1000pt

\newif\ifdtup

\relax
\newcommand{\be}{\begin{equation}}
\newcommand{\ee}{\end{equation}}
\newcommand{\bea}{\begin{eqnarray}}
\newcommand{\eea}{\end{eqnarray}}
\newcommand{\X}{{\vec X}}
\newcommand{\pro}{\partial}
\newcommand{\n}{\hat n}
\newcommand{\oneg}{\displaystyle\frac{1}{g}}

\newcommand{\D}{{\hat D}}

\newcommand{\A}{{\vec A}}
\newcommand{\valpha}{{\vec \alpha}}

\newcommand{\dfrac}{\displaystyle\frac}
\newcommand{\ba}{\begin{array}}
\newcommand{\ea}{\end{array}}

\newcommand{\nn}{\nonumber}
\begin{document}
\twocolumn[\hsize\textwidth\columnwidth\hsize\csname@twocolumnfalse%
\endcsname
\title  {Magnetic Confinement in QCD}
\bigskip

\author{Y. M. Cho$^{1, 2}$ and D. G. Pak$^{1, 3}$}

\address{
$^{1)}$Asia Pacific Center for Theoretical Physics\\
$^{2)}$Department of Physics, College of Natural Sciences, Seoul National University,
Seoul 151-742, Korea  \\
$^{3)}$Department of Theoretical Physics, Tashkent State University, Tashkent
700-095, Uzbekistan \\
{\scriptsize \bf ymcho@yongmin.snu.ac.kr,
dmipak@apctp.kaist.ac.kr} \\ \vskip 0.3cm
}
\maketitle

\begin{abstract}
We present a strong evidence for the magnetic confinement
in QCD by demonstrating that the one loop effective action of $SU(2)$
QCD induces a dynamical symmetry breaking thorugh 
the monopole condensation, which could induce the dual
Meissner effect and guarantee the confinement of color in the non-Abelian
gauge theory. The result is obtained by separating the topological degrees
which describes the non-Abelian monopoles from the dynamical degrees of
the potential, and integrating out all the dynamical degrees of QCD.

\vspace{0.3cm}
\end{abstract}

\narrowtext
\bigskip
                           ]
One of the most outstanding problems
in theoretical physics is the confinement problem in QCD. It has
long been argued that the monopole condensation could explain the
confinement of color through the dual Meissner effect \cite{nambu,cho1}.
Indeed, if one assumes the monopole condensation, one could easily argue that
the ensuing dual Meissner effect guarantees the confinement \cite{cho2,ezawa}.
But so far there has not been a satisfactory proof how the desired
monopole condensation could take place in QCD. In this direction, however,
there has been a remarkable progress in the lattice
simulation during the last decade.
In fact the recent numerical simulation has provided an unmistakable
evidence which supports the idea of the magnetic
confinement through the monopole
condensation \cite{kronfeld,stack}.
The purpose of this Letter is to establish
the magnetic confinement in QCD from the first principles.
{\it Utilizing the parameterization of the gauge potential
which emphasizes its topological character
we  establish the monopole condensation of vacuum in QCD
in one loop approximation,
after integrating out all the dynamical degrees of the non-Abelian potential.
The result confirms that the vacuum condensation naturally
generates the mass gap necessary for the dual Meissner effect}.
This demonstrates the existence of a dynamical symmetry breaking mechanism which
establishes the magnetic confinement in QCD. 

To prove the magnetic confinement it is instructive for us to
remember how the magnetic flux is confined in the superconductor
through the Meissner effect.  In the macroscopic Ginzburg-Landau description
of super conductivity the Meissner effect is triggered by the
mass term of the electromagnetic potential, which determines
the penetration (confinement) scale of the magnetic flux.
In the microscopic BCS description, this effective mass
is generated by the electron-pair (the Cooper pair) condensation.
This suggests that, for the confinement of the color electric flux,
one need the condensation of the monopoles.
Equivalently, in the dual Ginzburg-Landau description, one need the
dynamical generation of the effective mass for the
monopole potential. To demonstrate this one must first identify the
monopole potential, and separate it from the generic
QCD connection, in a gauge independent manner. This can be done
with the Abelian projection\cite{cho1,cho2},
 which provides us a natural reparameterization
of the non-Abelian connection in terms of
the dual potential of the maximal Abelian
subgroup $H$ of the gauge group $G$ and the gauge covariant vector
field of the remaining $G/H$ degrees.
 With this separation one can show that
the monopole condensation takes place after one
integrates out all the dynamical degrees of the non-Abelian gauge
potential. This strongly endorses the magnetic confinement in QCD.

Consider $SU(2)$ for simplicity.  A natural way to identify the
monopole potential is to introduce an isotriplet unit vector field
$\n$ which selects the ``Abelian'' direction at each space-time point, and to
decompose the connection into the Abelian part which leaves $\n$
invariant and the remaining part
which forms a covariant vector field \cite{cho1,cho2},
\bea
 & \vec{A}_\mu =A_\mu \n - \oneg \n\times\pro_\mu\n+\X_\mu\nonumber
         = \hat A_\mu + \X_\mu, \nn\\
 &  (\n^2 =1,~ \hat{n}\cdot\vec{X}_\mu=0),
\eea
where $
A_\mu = \n\cdot \vec A_\mu$
is the ``electric'' potential.
Notice that the Abelian projection $\hat A_\mu$ is precisely the connection which
leaves $\n$ invariant under the parallel transport,
\bea
\D_\mu \n = \pro_\mu \n + g {\hat A}_\mu \times \n = 0.
\eea
Under the infinitesimal gauge transformation
\bea
\delta \n = - \vec \alpha \times \n  \,,\,\,\,\,
\delta \A_\mu = \oneg  D_\mu \vec \alpha,
\eea
one has
\bea
&&\delta A_\mu = \oneg \n \cdot \pro_\mu \valpha,\,\,\,\
\delta \hat A_\mu = \oneg \D_\mu \valpha  ,  \nn \\
&&\hspace{1.2cm}\delta \X_\mu = - \valpha \times \X_\mu  .
\eea
Notice that $\hat A_\mu$ still describes an $SU(2)$ connection which
enjoys the full $SU(2)$ gauge degrees of freedom.
More importantly, $\hat{A}_\mu$ retains the full topological characteristics of the original non-Abelian potential.
Clearly the isolated singularities of $\hat{n}$ defines $\pi_2(S^2)$
which describes the non-Abelian monopoles.  Indeed $\hat A_\mu$
with $A_\mu =0$ and $\hat n= \hat r$ describes precisely
the Wu-Yang monopole \cite{wu,cho3}.  Besides, with the $S^3$
compactification of $R^3$, $\hat{n}$ characterizes the
Hopf invariant $\pi_3(S^2)\simeq\pi_3(S^3)$ which describes the topologically distinct vacuua
\cite{bpst,thooft}.

The above discussion tells that $\hat{A}_\mu$ has a dual
structure.
Indeed the field strength made of the restricted potential is decomposed as
\begin{eqnarray}
& \hat{F}_{\mu\nu} = (F_{\mu\nu}+ H_{\mu\nu})\hat{n}\mbox{,}\nonumber \\
& F_{\mu\nu} = \partial_\mu A_{\nu}-\partial_{\nu}A_\mu \mbox{,}\nonumber \\
& H_{\mu\nu} = -\frac{1}{g} \hat{n}\cdot(\partial_\mu
\hat{n}\times\partial_\nu\hat{n})
= \partial_\mu \tilde C_\nu-\partial_\nu \tilde C_\mu,
\end{eqnarray}
where $\tilde C_\mu$ is the ``magnetic'' potential
\cite{cho1,cho2}. So one can identify the non-Abelian
monopole potential by
\bea
\vec C_\mu= -\frac{1}{g}\hat n \times \partial_\mu\hat n ,
\eea
in terms of which the magnetic field is expressed by
\bea
\vec H_{\mu\nu}&=&\partial_\mu \vec C_\nu-\partial_\nu \vec C_\mu+ g \vec
C_\mu \times \vec C_\nu
=-\frac{1}{g} \partial_\mu\hat{n}\times\partial_\nu\hat{n} \nn\\
&=&H_{\mu\nu}\hat n.
\eea
Notice that the magnetic field has a remarkable structure
\bea
H_{\mu\alpha}H_{\alpha\beta}H_{\beta\nu} =-\dfrac{1}{2}
H^2_{\alpha\beta}H_{\mu\nu},
\eea
which will be very useful for us in the following.

With (1) one has
\bea
\vec{F}_{\mu\nu}&=&\hat F_{\mu \nu} + \D _\mu \X_\nu -
\D_\nu \X_\mu + g\X_\mu \times \X_\nu,
\eea
so that the Yang-Mills Lagrangian is expressed as
\bea
{\cal L} = &-&\dfrac{1}{4} \vec F^2_{\mu \nu }
=-\dfrac{1}{4}
{\hat F}_{\mu\nu}^2 -\dfrac{g}{2} {\hat F}_{\mu\nu} \cdot (\X_\mu \times \X_\nu)  \nn \\
 &-&\dfrac{1}{4}(\D_\mu\X_\nu-\D_\nu\X_\mu)^2-\dfrac{g^2}{4} (\X_\mu \times \X_\nu)^2. 
\eea
This shows that the Yang-Mills theory can be viewed as
the restricted gauge theory made of the Abelian projection,
which has an additional gauge covariant charged vector field
(the valence gluons) as its source \cite{cho1,cho2}. Obviously the theory
is invariant
under the gauge transformation (3) of the active type. But notice that
it is also invariant under the following gauge transformation
of the passive type,
\bea
\delta\hat{n}=0,~~~~\delta\vec{A}_{\mu}=\dfrac{1}{g}D_{\mu}\vec{\alpha},
\eea
under which one has
\bea
&\delta A_{\mu}=\dfrac{1}{g}\hat{n}\cdot D_{\mu}\vec{\alpha},~~~~~
\delta \vec{C}_\mu=0 , \nonumber\\
&\delta\vec{X}_{\mu}=\dfrac{1}{g}[D_{\mu}\vec{\alpha}
 -(\hat{n}\cdot D_{\mu}\vec{\alpha})\hat{n}]  .
\eea
This gauge invariance of the passive type will be useful in the following.

With this preparation
we will now show that the effective theory of QCD, which
one obtains after integrating out all the dynamical degrees,
can be written in one loop
approximation as
\bea
&~~{\cal L}_{eff} = -\dfrac{Z}{4} \vec{H}^2_{\mu\nu}, \nn\\
&Z=1+\dfrac{22}{3}\dfrac{g^2}{(4\pi)^2}\Big(\ln\dfrac{gH}{\mu^2}-c\Big),
\eea
where $H=\sqrt{\vec{H}_{\mu\nu}^2}$,  $\mu$ is a dimensional parameter, and
$c$ is a finite constant.
This is our main result, which provides
the desired
monopole condensation of the vacuum.

To derive the effective action consider the generating functional of (10)
\bea
W[J_\mu, {\vec J}_\mu] &=&
\int {\cal D}A_\mu {\cal D} \X_\mu  \exp [ i \int
(- \dfrac{1}{4} {\vec F}_{\mu\nu}^2  \nn \\
&+& A_\mu J_\mu + \X_\mu \cdot {\vec J}_\mu) d^4 x  ] .
\eea
We have to perform the functional integral
with a proper choice of a gauge, leaving $\vec C_\mu$ as a background.
With (11) we choose the gauge fixing condition
\bea
{\vec F}&=& \hat D_\mu (A_\mu \hat n + \vec X_\mu) =0\nn\\
{\cal L}_{gf}= &-& \dfrac{1}{2\xi}
\left[(\partial_\mu A_\mu)^2 + ({\hat D}_\mu \X_\mu)^2\right].
\eea
With this the generating functional takes the form,
\bea
W[J_\mu,{\vec J}_\mu]&=&\int {\cal D} A_\mu {\cal D}
\X_\mu {\cal D} \vec{c} {\cal D}\vec{c}^{~*}
\exp \{{~i \int[-\dfrac {1}{4}{\vec F}_{\mu \nu}^2} \nn \\
&+&\vec{c}^{~*}\hat{D}_\mu D_\mu\vec{c}
-\frac{1}{2\xi}(\partial_\mu A_\mu)^2-\frac{1}{2\xi}
(\hat{D}_\mu\vec{X}_\mu)^2\nn\\
&+& A_\mu J_\mu+\X_\mu \cdot \vec J_\mu]d^4x\},
\eea
where $\vec c$ and ${\vec c}^{~*}$ are the ghost fields. In one loop
approximation the $A_\mu$ integration becomes trivial,
and the $\X_\mu$ and ghost integrations result in the
following functional determinants (with $\xi=1$),
\bea
&{\rm Det}^{-\frac{1}{2}} K_{\mu \nu}^{ab}\simeq
{\rm Det}^{-\frac{1}{2}}[g_{\mu \nu}
 (\tilde D \tilde D)^{ab}
- 2gH_{\mu \nu}\epsilon^{abc} n^c], \nn\\
&{\rm Det} M^{ab}_{FP} \simeq {\rm Det} (\tilde{D} \tilde{D})^{ab},
\eea
where $\tilde D_\mu$ is defined with only $\vec C_\mu$.
One can simplify the determinant $K$ 
using the relation (8),
\bea
\ln {\rm Det}^{-\frac{1}{2}} K& =&-\ln {\rm Det}(\tilde{D}\tilde{D})^{ab}\nn\\
&-&\frac12\ln {\rm Det}[(\tilde{D}\tilde{D})^{ab}
+i\sqrt{2}gH\epsilon^{abc}n^c]\nn\\
&-&\frac12\ln {\rm Det}[(\tilde{D}\tilde{D})^{ab}-i\sqrt{2}gH\epsilon^{abc}n^c].
\eea
With this the one loop contribution of the functional
determinants to the effective action can be written as
\bea
\Delta S=i\ln {\rm Det}(\tilde{D}^{2} +\sqrt{2}gH)
               (\tilde{D}^{2} -\sqrt{2}gH),
\eea
where now $\tilde{D}_\mu$ acquires the following Abelian form,
\bea
\tilde{D}_\mu =\partial_\mu + ig\tilde{C}_\mu .\nn
\eea
Notice that the reason for this simplification is precisely because
$\vec{C}_\mu$ originates from the Abelian projection.
With this one can use the heat kernel method and the zeta
function regularization
to calculate the functional determinant.
For the covariantly constant $\vec H_{\mu\nu}$ we find
\bea
\Delta{\cal L} = \zeta'_{+}(0)+\zeta'_{-}(0),
\eea
where $\zeta_{\pm}$ is a generalized zeta function given by
\bea
\zeta_{\pm}(s)&=&\frac{\pm i}{16\pi^2\Gamma(s)}
\int^{\infty}_{0}(\pm i\tau)^{s-3}
\frac{gH\tau/\sqrt{2}\mu^2}{\sin[ gH(\tau\pm i\varepsilon)/\sqrt{2}\mu^2]}\nn\\
&&\times\exp[\pm\sqrt{2}igH(\tau\pm i\varepsilon)/\mu^2]d\tau.
\eea
From this we finally obtain (with the modified minimal subtraction)
\bea
&{\cal L}_{eff}=-\dfrac{1}{4}H^2 -\dfrac{11g^2}{96\pi^2}H^2(\ln
\dfrac{gH}{\mu^2}-c ), \nn\\
&c=1+\dfrac{15}{22}\ln2+\dfrac{12}{11}\zeta'(-1)=1.2921409..... ,
\eea
where now $\zeta(s)$ is the Riemann's zeta function.
This completes the derivation of the effective Lagrangian (13).

Clearly the effective action provides the following non-trivial
effective potential
\bea
V&=&\frac{g^2}{4}(\vec{C}_\mu\times\vec{C}_\nu)^2\Big \{ 1 \nn \\
&+&\frac{22}{3}\frac{g^2}{(4\pi)^2}
\Big[ \ln \frac{g^2[(\vec{C}_\mu\times\vec{C}_\nu)^2]^{1/2}}
{\mu^2}-c\Big]\Big\},
\eea
which generates the desired magnetic condensation of the vacuum,
\bea
<H>=\frac{\mu^2}{g} \exp\Big(-\frac{24\pi^2}{11g^2}+ c-\frac12\Big).
\eea
The vacuum generates an ``effective mass'' for $\vec C_\mu$,
\bea
m^2=\frac{11g^4}{96\pi^2}\Big<\frac{(\vec{C}_\mu\times
\vec{H}_{\mu\nu})^2}{H^2}\Big>,
\eea
which demonstrates that the magnetic condensation indeed generates the
mass gap necessary for the dual Meissner effect. Obviously the mass scale
sets the confinement scale.

To check the consistency of our result with the perturbative QCD
we now discuss the running coupling and the renormalization.
For this we define the running coupling $\bar g$ by
\bea
\frac{\partial^2V}{\partial H^2}\Big|_{H=\bar H} =\frac{1}{2}\frac{g^2}{ \bar g^2}.
\eea
So with $g \bar H = \bar\mu^2 \exp(c-3/2)$
we obtain the following $\beta$-function,
\bea
\frac{1}{\bar g^2} =
\frac{1}{g^2}+\frac{11}{12\pi^2}\ln\frac{\bar\mu}{\mu},~~~
\beta(\bar\mu)=-\frac{11}{24\pi^2} \bar g^3~,
\eea
which exactly coincides with the well-known result \cite {gross}.
In terms of the running coupling the renormalized potential is given by
\bea
V_{\rm ren}=\frac14 H^2\Big[1+\frac{22}{3}\frac{\bar g^2}{(4\pi)^2}
(\ln\frac{\bar g H}{\bar\mu^2}-c)\Big],
\eea
and the Callan-Symanzik equation
\bea
\Big(\bar\mu\frac{\partial}{\partial \bar\mu}+\beta\frac{\partial}{\partial \bar g}
-\gamma\vec{C}_\mu\frac{\partial}{\partial\vec{C}_\mu} \Big)V_{\rm ren}=0,
\eea
gives the following anomalous dimension for $\vec C_\mu$,
\bea
\gamma=-\frac{11}{24\pi^2}\bar g^2.
\eea
This should be compared with that of the gluon field in perturbative QCD,
$\gamma(\vec{A}_\mu)=5\bar g^2/24\pi^2$ for $SU(2)$.

Notice that with the vacuum condensation
the effective Lagrangian can be approximated
as
\bea
{\cal L}_{eff}&\simeq&-\frac14\vec{H}_{\mu\nu}^2
                         -\frac12m^2\vec{C}_\mu^2\nn\\
&=&-\frac{m^2}{2g^2}(\partial_\mu \hat{n})^2-\frac{1}{4g^2}(\partial_\mu \hat{n}
\times\partial_\nu \hat{n})^2.
\eea
This is nothing but the Skyrme-Faddeev Lagrangian which allows the
topological knot solitons as the classical solutions \cite {faddeev}.
It is truly remarkable that the Skyrme-Faddeev Lagrangian can be
derived from the effective Lagrangian of QCD.
Our analysis establishes the deep connection that exists
between the generalized non-linear sigma model of Skyrme-Faddeev type
and QCD.

We conclude with the following remarks: \\
1) One might question (legitimately)
the validity of the one loop approximation,
since in the infra-red limit the non-perturbative effect
is supposed to play the essential role
in QCD. Our attitude on this issue is that {\it QCD can be viewed as the
perturbative extension of the topological field theory described
by the restricted QCD, so that the non-perturbative
effect in the low energy limit can effectively be represented by
the topological structure of the restricted gauge
theory}. This is reasonable,
because the large scale structure of the monopole topology
naturally describes the long range behavior
of the theory. In fact one can show that it is the restricted connection
that contributes to the Wilson loop integral,
which provides the confinement criterion in QCD
\cite{cho4}.
So we believe that our classical monopole
background automatically
takes care of the essential feature of the non-perturbative effect,
which should make the one loop approximation reliable. \\
2) Our vacuum looks very much like the old ``Savvidy vacuum'' \cite{savv}.
But notice that, unlike the Savvidy vacuum, ours is stable.
In this connection we emphasize that one must be very careful when one
calculates the functional determinant (19). Naively Det K
contains negative eigenvalues whose eigenfunctions become tachyonic,
which cause an infra-red divergence. So one must exclude these 
unphysical tachyonic modes with a proper infra-red regularization,
when one calculates the functional determinant. Only with the
exclusion of the unphysical modes one can obtain a consistent 
theory of QCD \cite{cho5}. \\
3) There have been two competing proposals for the correct mechanism
of the confinement in QCD, the one emphasizing the role of the instantons and
the other emphasizing that of the monopoles. Our analysis strongly
favors the monopoles as the physical
source for the confinement. It
provides the correct dynamical symmetry breaking, and generates the mass
gap necessary for the confinement in the infra-red limit of QCD.

It must be clear from our analysis that
the existence of the magnetic condensation is a generic
feature of the non-Abelian gauge theory.
A more detailed discussion, including vacuum stability and the the
generalization of our result to $SU(3)$,
will be presented in a forthcoming paper \cite{cho5}.

One of the authors (YMC) thanks Professor C. N. Yang for
the fruitful discussions and the encouragements. The other (DGP)
thanks him for the fellowship at Asia Pacific Center for Theoretical
Physics, and appreciates Haewon Lee
for numerous discussions.
The work is supported in part by Korean Science and Engineering
Foundation, and by the BK21 project of Ministry of Education.


\begin{thebibliography}{99}
\bibitem{nambu}Y. Nambu, Phys. Rev. {\bf D10}, 4262 (1974);
S. Mandelstam, Phys. Rep. {\bf 23C}, 245 (1976);
A. Polyakov, Nucl. Phys. {\bf B120}, 429 (1977);
G. 't Hooft, Nucl. Phys. {\bf B190}, 455 (1981).
\bibitem{cho1}Y. M. Cho, Phys. Rev. {\bf D21}, 1080 (1980); JKPS, {\bf17}, 266 (1984).
\bibitem{cho2}Y. M. Cho, Phys. Rev. Lett. {\bf 46}, 302 (1981); Phys. Rev. {\bf D23}, 2415 (1981).
\bibitem{ezawa}Z. Ezawa and A. Iwazaki, Phys. Rev. {\bf D25}, 2681 (1982);
T. Suzuki, Prog. Theor. Phys. {\bf 80}, 929 (1988);
H. Sugamura, S. Sasaki, and H. Toki, Nucl. Phys. {\bf B435}, 207 (1995);
S. Shabanov, Mod. Phys. Lett. {\bf A11}, 1081 (1996);
R. Brower, K. Orginos, and C. I. Tan, Phys. Rev. {\bf D55}, 6313 (1997);
K. Kondo, Phys. Rev. {\bf D57}, 7467 (1998); {\bf D58}, 105016 (1998).
\bibitem{kronfeld}A. Kronfeld, G. Schierholz, and U. Wiese, Nucl. Phys. {\bf B293}, 461 (1987);
T. Suzuki and I. Yotsuyanagi, Phys. Rev. {\bf D42}, 4257 (1990).
\bibitem{stack}J. Stack, S. Neiman, and R. Wensley, Phys. Rev. {\bf D50}, 3399 (1994);
H. Shiba and T. Suzuki, Phys. Lett. {\bf B333}, 461 (1994);
G. Bali, V. Bornyakov, M. M\"{u}ller-Preussker, and K. Schilling, Phys. Rev. {\bf D54}, 2863 (1996).
\bibitem{wu}T. T. Wu and C. N. Yang, Phys. Rev. {\bf D12}, 3845 (1975).
\bibitem{cho3} Y. M. Cho, Phys. Rev. Lett. {\bf 44}, 1115 (1980); Phys. Lett.
{\bf B115}, 125 (1982).
\bibitem{bpst}A. Belavin, A. Polyakov, A. Schwartz, and Y. Tyupkin, Phys. Lett. {\bf 59B}, 85 (1975); Y. M. Cho,
Phys. Lett. {\bf B81}, 25 (1979).
\bibitem{thooft} G. 't Hooft, Phys. Rev. Lett. {\bf 37}, 8 (1976); R. Jackiw and C. Rebbi, Phys. Rev. Lett.
{\bf 37}, 172 (1976); C. Callan, R. Dashen, and D. G. Gross, Phys. Lett. {\bf 63B}, 334 (1976).
\bibitem{gross} D. Gross and F. Wilczek, Phys. Rev. Lett. {\bf 26}, 1343 (1973);
H. Politzer, Phys. Rev. Lett. {\bf 26}, 1346 (1973).
\bibitem{faddeev} L. Faddeev and A. Niemi, Nature {\bf 387}, 58 (1997);
R. Battye and P. Sutcliffe, Phys. Rev. Lett. {\bf 81}, 4798 (1998);
L. Faddeev and A. Niemi, Phys. Rev. Lett. {\bf 82}, 1624 (1999);
Phys. Lett. {\bf B449}, 214 (1999).
\bibitem{cho4}Y. M. Cho, hep-th/9905127, submitted to Phys. Rev. {\bf D}.
\bibitem{savv} G. Savvidy, Phys. Lett. {\bf B71}, 133 (1977);
N. Nielsen and P. Olesen, Nucl. Phys. {\bf B144}, 485 (1978).
\bibitem{cho5} Y. M. Cho and D. G. Pak, hep-th/0006051, in {\it Proceedings
of TMU-Yale Symposium on Dynamics of Gauge Fields}, edited by T. Appelquest and
H. Minakata (Universal Academy Press, Tokyo) 1999.
\end{thebibliography}
\end{document}